**DNA origami nanotechnology for building artificial dynamic systems**

Na Liu

Na Liu, Max Planck Institute for Intelligent Systems, and Kirchhoff Institute for Physics, University of Heidelberg, Germany; na.liu@kip.uni-heidelberg.de

A fundamental design rule that nature has developed for biological machines is the intimate correlation between motion and function. One class of biological machines is molecular motors in living cells, which directly convert chemical energy into mechanical work. They coexist in every eukaryotic cell, but differ in their types of motion, the filaments they bind to, the cargos they carry, as well as the work they perform. Such natural structures offer inspiration and blueprints for constructing DNA-assembled artificial systems, which mimic their functionality. In this article, we describe two groups of cytoskeletal molecular motors, linear and rotary motors. We discuss how their artificial analogs can be built using DNA origami technology. Finally, we summarize ongoing research directions and conclude that DNA origami has a bright future ahead.

**Introduction**

Through evolution, nature has found the best and the most efficient designs for machinery, which provides inspiration for numerous technological developments. We constantly look for solutions to various challenges by learning from nature. One of the most important inventions in the history of humankind is airplanes, which take the shape of birds.[1] Another example is the beautiful wings of butterflies, whose intrinsic structures have inspired new concepts in displays,[2] fabrics, and cosmetics.[3]

Looking further down to the molecular level, nature is also extremely efficient in constructing biological machines.[4] One class of biological machines is molecular motors in living cells, which directly convert chemical energy into mechanical work.[5] Such motor proteins, including kinesin,[6] dynein,[7] and myosin,[8], are all powered by the hydrolysis of adenosine triphosphate (ATP), known as the fuel of life. Even though many unsolved questions remain towards the complete understanding of their working mechanisms, our current knowledge about these biological building blocks represents a precious treasure for nanotechnologists. Their optimized structures and operating mechanisms provide proof of feasibility to use nanotechnology for the realization of energy transfer, material delivery, and information processing, as well as other processes on the nanoscale.[9] In turn, progress in the development of nanotechnology will allow engineering molecular entities for specific functions and needs in the future.

**DNA origami nanotechnology**

Among various biomimetic approaches, DNA origami nanotechnology is one of the most powerful tools to build artificial nanosystems entirely from the bottom up.[10,11] The unique specificity of DNA interactions and our ability to code DNA sequences and to chemically functionalize DNA make it ideal for controlling the self-assembly of nanoscale components with high fidelity. In the 1980s, N. Seeman ushered in DNA as a construction material into the nanoscale world.[12] In 2006, a crucial breakthrough in the field of DNA nanotechnology was achieved with the concept of DNA origami, developed by P. Rothemund.[10]

DNA origami involves the folding of a long scaffold DNA strand by hundreds of short staple strands into nanostructures with nearly arbitrary shapes. As each staple strand possesses a unique sequence and its position in the formed structure is deterministic, DNA origami can serve as "molecular pegboards" for self-assembly of nanoscale elements with significant geometric and topological complexity.[13,14] Most importantly, this technology enables dynamic functionality of the assembled structures upon a variety of external inputs.[15–18] This sets a solid foundation to realize DNA-based programmable nanomachinery.

DNA origami nanotechnologists are lucky. On the one hand, they possess a unique nanoscale engineering tool to self-assemble artificial nanodevices, which can perform controlled motion in response to specific stimuli with programmability and addressability on the molecular level. On the other hand, they never need to worry about a lack of ideas—biological machinery offers inspiration and blueprints for them to follow and advance the state-of-the-art in nanotechnology. As R. Feynman stated, "There is plenty of room at the bottom."[19]

**Artificial nanosystems enabled by DNA origami nanotechnology**

A cell is a biological factory, containing myriad molecular motors, working together to accomplish complex biological functions.[20] Molecular motors can be classified into two groups, linear motors and rotary motors.[9] Kinesin, dynein, and myosin are linear motors. They serve as molecular transporters to deliver vesicles, molecules, and organelles throughout a cell.[21] Due to the crowded nature of cytosols, the intracellular fluids and the relatively substantial size of a cell, transporting materials from one location to another through diffusion is too slow and inefficient. Molecular transporters circumvent this hurdle by performing directed movement fueled by ATP hydrolysis along polarized cytoskeletal filaments. Most kinesins unidirectionally transport cargos toward the plus-ends of the microtubules, whereas all dyneins walk toward the minus-ends.[22] Myosins move along actin filaments, playing an essential role in muscle contraction and the transport of secretory vesicles in yeasts.[23] $F_OF_1$-ATP synthase is a membrane-bound rotary motor found in all organisms. It produces ATP powered by the electrochemical proton gradient across the membrane, and its subunits can carry out 360° rotary motion.[24]

Construction of biomimetic nanomachines that carry out controlled motion and specific tasks is an everlasting goal of nanotechnology. For instance, exciting progress has been witnessed in the realization of DNA-assembled artificial systems, which mimic the functionality of molecular motors. As DNA origami nanotechnologists, we are clearly aware of the fact that at this stage it is impossible to create any biomimetic systems that bear the same complexity and sophistication of natural systems. Nevertheless, attempts along this direction will help us gain general knowledge related to self-assembly in a systemic way. In addition, the continuous endeavors will provide profound insights into how machines, and generally how functional systems, have to be built on the nanoscale. This is also a prerequisite for the coming era of bionanotechnology, leading to the eventual convergence of biology and nanotechnology.

*Walking*

Kinesin-1 is a microtubule-based molecular motor (**Figure 1**a).[20] It is a heterotetramer, which consists of two heavy chains, each with a motor domain, a coiled tail, and two light chains that bind cellular cargos. The walking of kinesin-1 is bipedal with the two heads binding alternatively to a microtubule. For each 8-nm step, it hydrolyzes one ATP molecule.[6,25] Kinesin-1 walks along a microtubule with a maximum speed of ~800 nm/s and can move continuously for up to several micrometers without detaching.[6,25]

The remarkable function of kinesin-1 has greatly inspired scientists to build artificial analogues using DNA origami nanotechnology. For instance,

Gu et al., demonstrated a proximity-based programmable DNA nanoscale assembly line, shown in Figure 1b.[26] A DNA origami tile served as a platform, on which a DNA walker could walk and collect nanoparticle cargos from DNA switches located at three specific locations. Through a sequence of strand additions, the switches physically placed the nanoparticles either within or outside the DNA origami. The DNA origami was functionalized with toeholds for the DNA walker. Through strand displacement reactions, the DNA walker could walk to the switches and collect cargos in a fully programmable manner. The progression of the walker was monitored using atomic force microscopy. Zhou et al. demonstrated a gold nanorod (AuNR) itself worked as a walker, which could move 7 nm each step linearly on a DNA origami track (Figure 1c).[27] The walker AuNR was functionalized with foot strands. To enable robust binding, at each station the walker stepped on two rows of neighboring footholds extended from the DNA origami. The stepwise walking was powered by DNA hybridization and activated upon addition of respective blocking and removal DNA strands. The optical tracking of the stepwise walking process was enabled by the near-field interactions of the walker and the stator AuNR immobilized on the bottom surface of the origami, through circular dichroism spectral changes in real time.

Thubagere et al. recently demonstrated a DNA robot that could collectively perform a cargo sorting task on a DNA origami surface by transporting fluorescent molecules of different colors from initially unordered locations to separated destinations (Figure 1d).[28] The authors designed an algorithm such that the DNA robot could pick up multiple cargos of two types and delivered each type to a predesignated location until all cargo molecules were sorted. Remarkably, the DNA robot could perform a random walk without any energy supply and carried out approximately 300 steps while sorting the cargos. This work shed light on the development of multitasking nanoscale robots and will enable profound applications in autonomous chemical synthesis and programmable therapeutics.

*Sliding*

There are also molecular motors that cause cytoskeletal filaments to slide against each other and generate forces, leading to muscle contraction, ciliary beating, and cell division.[20] For instance, kinesin-5 is essential for mitosis in most organisms. It is a homotetramer with two motor domains at each end as shown in **Figure 2**a. Kinesin-5 cross-links antiparallel microtubules and slides apart duplicated centrosomes during the formation of the mitotic spindle by consuming chemical energy from ATP hydrolysis. Different from other kinesins, kinesin-5 can simultaneously interact with two microtubules. Therefore, the microtubule is both the cargo and track for kinesin-5.[29] The speed of kinesin-5 is approximately 20–100 nm/s, much slower than that of kinesin-1.[30,31] When sliding, a large mechanical force on the order of several piconewtons per molecule can be generated.

Urban et al. recently demonstrated a nanoscopic sliding system, in which gold nanocrystals (AuNCs) could mediate coordinated sliding of two antiparallel DNA origami filaments by addition of DNA strands (see Figure 2b).[32] Two 14-helix DNA origami filaments were cross-linked by two AuNCs. Ten rows of footholds evenly separated by 7 nm were extended from each origami filament. The foothold rows were reversely positioned along the two filaments, whose polarities were defined using "+" and "–." The two AuNCs were bound in between the filaments with the same combination of four foothold rows (i.e., two rows from each filament). This resembles the homotetramer structure of the kinesin-5 protein, which comprises four motor domains, two on each end to interact with the microtubule tracks. The two AuNCs could be driven simultaneously through toehold-mediated strand displacement reactions by the same sets of DNA fuels. To optically monitor the sliding process in real time, two fluorophores were tethered at one end of the system, correlating subtle spatial changes with optical response changes.

Sliding motion has also been achieved using DNA origami structures by imitating rotaxane-based molecular machines. A rotaxane is a mechanically interlocked molecular architecture, which consists of a dumbbell-shaped molecule threaded through a macrocycle.[33] The macrocycle can slide the axis of the dumbbell from one end to another. Marras et al. realized a crank-slider origami system, which consisted of three hinges and one slider joined to combine rotational and linear motion.[34] Thermal fluctuations caused the joint to extend and contract along the linear motion path.

List et al. later demonstrated a more controllable system, as shown in **Figure 2**c.[35] An origami ring was threaded by an origami dumbbell. Through toehold-mediated strand displacement reactions, the connection between these two origami units could be locked or unlocked. As a result, the origami ring could be switched between a localized state and a mobile state by sliding along the origami dumbbell axis over several hundreds of nanometers.

*Rotation*

ATP is the most commonly used energy currency in cells for all organisms.[9,36] $F_OF_1$-ATP synthase is an enzyme that creates ATP. It consists of two distinct units, $F_O$ and $F_1$, as shown in **Figure 3**a. The $F_O$ moiety is embedded in the cell membrane, whereas the $F_1$ moiety protrudes from the membrane. They are elastically coupled by a common shaft $\gamma$ and stabilized by a peripheral stator-stalk. The activity of ATP synthase is reversible.[9] When running in the forward direction, ATP is produced outside the membrane. When running backward, it hydrolyzes ATP into ADP (adenosine diphosphate) and phosphate, pumping protons across the membrane. It has been experimentally observed that at low ATP concentrations, the central $\gamma$-subunit rotates in discrete 120° steps, whereas at high ATP concentrations the rotation becomes continuous and saturates at ~130 revolutions per second.[37,38]

Rotary devices have been constructed using DNA origami based on different mechanisms. Nevertheless, most systems only carried out simple switching between several predesignated configurations. Kuzyk et al. demonstrated a reconfigurable three-dimensional (3D) metamolecule scheme, as shown in Figure 3b.[39] Two origami bundles are linked together by a scaffold strand to form a reconfigurable DNA origami template, which was utilized to host two AuNRs by DNA hybridization. Upon interacting with light, the excited plasmons in the two AuNRs were strongly coupled due to their close proximity, leading to a plasmonic chiral system. The switching was enabled by the two DNA locks that were extended from the sides of the DNA origami bundles. By adding specifically designed DNA fuel strands, toehold-mediated strand displacement reactions were triggered, and these DNA locks could be opened or closed accordingly. As a result, the conformation of the 3D metamolecule could be switched between different states, thus giving rise to circular dichroism spectral changes.

Ketterer et al. reported a nanoscale rotary system, which was built from tight-fitting components self-assembled using multilayer origami, as shown in Figure 3c.[40] The rotary apparatus consisted of a rotor unit and two clamp elements, which formed an axle bearing. In order to understand the rotary mechanism by optical spectroscopy, the authors extended the crank lever of the rotor unit with a 430 nm DNA origami filament. At its tip, six reporter fluorophores were labeled. The apparatus was attached to a surface and single-particle fluorescence microscopy characterizations were carried out. It was demonstrated that the rotor could perform random rotation along a circle with one end attached to the surface. Kopperger et al. reported a self-assembled nanoscale robotic arm controlled by electric fields (see Figure 3d).[41] The system was composed of an origami plate with an integrated 25-nm-long origami bundle. The two parts were connected via a flexible joint created by two adjacent scaffold crossovers. To enable direct observation of the arm's motion by fluorescence microscopy, the arm was linearly extended up to several hundred nanometers via a shape-complementary connector structure labeled with fluorophores. An electrophoretic chamber with two perpendicular fluidic channels was utilized for the electrical control experiments. The DNA nanostructures were immobilized in the center of the chamber and experienced a superposition of the fields generated from the electrode pairs. Arbitrary control of the pointing direction of the arm could be achieved by applying a voltage. Electrical actuation of the arms was directly observed with a charge-coupled device using total internal reflection fluorescence microscopy. Remarkably, the arm could be switched between arbitrary positions on the platform within milliseconds.

**Summary**

One of the most exciting objectives in the field of DNA nanotechnology is to accomplish advanced artificial nanofactories, in which the spatial arrangements of different components and most functions are enabled by DNA. At this stage, DNA origami nanotechnologists have already learned a great deal from biological machines and gained profound knowledge about how DNA machines can be built on the nanoscale. There is certainly a long path ahead to fully mimic the functional characteristics of biological machines, which have been optimized through billions of years of evolution. However, with DNA origami nanotechnology as a key at hand for nanoscale engineering and the advent of deeper insights, we believe construction of fully functional nanomachines that can carry out complicated tasks will not only exist in our imagination.

From a broader perspective, 13 years of research efforts after the breakthrough work of P. Rothemund on DNA origami have fostered and brought forward many inspiring research directions in DNA nanotechnology. For nanophotonics, DNA origami has been used to assemble a variety of inorganic elements, including metal and semiconductor nanoparticles as well as fluorophores into functional structures.[42] This not only provides alternative approaches to build practical devices such as solar cells, light-emitting diodes, and photonic circuits, but also offers unique platforms to manipulate light–matter interactions with nanoscale accuracy.

For synthetic biology, DNA origami has been used to assemble proteins,[43] given the unique programmability of DNA. Understanding protein structures was actually the original vision of N. Seeman, one of the founders of DNA nanotechnology. DNA origami can serve as templates to organize proteins in ordered arrays as well as to create controllable enzyme cascades for enhancing enzymatic activities.[44]

For biophysics, DNA origami has been used in the development of super resolution fluorescence techniques, such as DNA points accumulation for imaging nanoscale topography (DNA-PAINT).[45] DNA-PAINT relies on transient binding and unbinding of oligonucleotide strands labeled with fluorophores to specific positions of DNA origami. It enables simultaneous imaging of microtubules, mitochondria, and the Golgi apparatus and peroxisomes inside fixed cells.

For biomedical applications, DNA origami also holds great promise, especially in robotic drug delivery.[46] DNA origami can be fabricated into different shapes and sizes, be functionalized with different ligands, and, most importantly, be dynamically controlled. As a result, drug delivery with DNA nanotechnology can become fully programmable.

Due to its multidisciplinary nature, the field of DNA origami nanotechnology uniquely gathers scientists from physics, chemistry, biology, engineering, materials science, computer science, and others. Thanks to their tremendous endeavors, this field has already departed from its infancy and stepped into an exciting phase, where research ideas are being vigorously transferred into functional devices. The cheerful DNA origami nanotechnologists always look forward to new challenges and opportunities. We believe that by its 20th anniversary, a new spectrum of inspiring research will be established, and DNA origami will already be utilized for real-world applications.

**Acknowledgments**

We thank the financial support from the Volkswagen foundation and a European Research Council (ERC Dynamic Nano) grant.

46. S. Li, Q. Jiang, S. Liu, Y. Zhang, Y. Tian, C. Song, J. Wang, Y. Zou, G.J. Anderson, J.-Y. Han, *Nat. Biotechnol.* **36**, 258 (2018).

**Figure Captions**

**Figure 1.** (a) Schematic of kinesin-1 and kinesin-2 molecular motors, which perform stepwise walking along microtubules for organelle transport. Reprinted with permission from Reference 20. © 2008 Macmillan Publishing. The microtubule is made of polymerized α- and β-tubulin dimers characterized by light and dark green colors. (b) A DNA walker moves and transfers cargos on a DNA origami tile.[26] The DNA origami tile serves as a platform, on which a DNA walker walks and collects nanoparticle cargos from DNA switches located at three specific locations. A-k binds Fk to the origami and FA-k is a fuel strand that removes A-k, undoing the corresponding binding. Foot-binding sites on the origami are labelled such that in its nth binding to the origami, Fk binds to site kn. Reprinted with permission from Reference 26. © 2010 Springer Nature. (c) Gold nanorod (yellow) walks stepwise on a DNA origami track (grey rectangular platform) upon addition of specifically designed DNA strands. The detailed DNA sequences are provided in the insets. Each step is 7 nm. A: adenine; T: thymine; G: guanine; C: cytosine. During this process, it optically interacts with another gold nanorod (red), leading to circular dichroism spectral changes in real time.[27] Reprinted with permission from Reference 27. © 2015 Springer Nature. (d) Two DNA robots that perform a cargo sorting task on a DNA origami surface, transporting fluorescence molecules (green and yellow) to separate destinations based on a well-designed algorithm. Reprinted with permission from Reference 28. © 2017 AAAS.

**Figure 2.** (a) Schematic of kinesin-5, a bipolar homotetrameric motor protein, uses pairs of dimeric motor domains to cross-link and slide apart antiparallel microtubules. Reprinted with permission from Reference 20. © 2008 Macmillan Publishing. The microtubule is made of polymerized α- and β-tubulin dimers characterized by light and dark green colors. The two motor domains are at the two ends of the protein (purple). (b) Two gold nanocrystals (yellow) cross-link and slide two DNA origami filaments in antiparallel.[32] Upon addition of blocking strands 4 and 8 and removal strands $\bar{2}$ and $\bar{6}$, toehold-mediated strand displacement reactions take place. Rows 4 and 8 are blocked and the gold nanocrystals are released from these rows. Meanwhile, rows 2 and 6 are activated to bind the gold nanocrystals. As a result, the two gold crystals slide the filaments relative to one another for one step in a cooperative manner, introducing a 14 nm displacement. The numbers on the origami filaments represent the locations of the footholds. Different footholds are illustrated using different colors. D: donor; A: acceptor. (c) Fuel/antifuel mechanism used to switch the connection between the two units in a DNA origami rotaxane through toehold mediated strand displacement reactions, enabling long-range sliding

movement. Reprinted with permission from Reference 35. © 2016 Springer Nature. The red and blue stars represent fluorophores. The two units are origami dumbbell and ring, respectively.

**Figure 3.** (a) Schematic of the $F_OF_1$-ATP synthase complex. ATP synthase consists of two regions $F_O$ and $F_1$. $F_O$ causes rotation of $F_1$ and is made of c-ring and subunits. $F_1$ is made of $\alpha$, $\beta$, $\gamma$, $\delta$ subunits. The $F_O$ moiety is embedded in the cell membrane (purple), whereas the $F_1$ moiety protrudes from the membrane. Reprinted with permission from Reference 36. © 2011 AAAS. (b) Reconfigurable plasmonic metamolecule that is switched between predesignated configurations.[39] Two gold nanorods are hosted on a switchable DNA origami template consisting of two connected bundles, which subtends a tunable angle. The relative angle between the rods and therefore the handedness of the 3D chiral nanostructure can be actively controlled by two DNA locks, which are extended from the sides of the DNA origami template. Specifically designed DNA strands work as fuel to drive the plasmonic nanostructure to desired states with distinct 3D conformations by altering the relative angle between the two DNA bundles and hence the rods. The functional segments of the DNA fuels are illustrated using different colors. (c) Schematic of a nanoscale rotary apparatus formed from tight-fitting DNA origami components. The bottom-right image is the top view of the system. Reprinted with permission from Reference 40. © 2016 AAAS. (d) Self-assembled nanoscale robotic arm (blue bundle) on an origami plate (grey rectangle) controlled by electric fields for rotation. Reprinted with permission from Reference 41. © 2018 AAAS.

**Na Liu** is a professor at the Kirchhoff Institute for Physics at the University of Heidelberg, Germany. She received her PhD degree in physics at the University of Stuttgart, Germany. Her research focuses on developing sophisticated and smart optical nanosystems for answering structural biology questions as well as catalytic chemistry questions in local environments. Liu can be reached by email at na.liu@kip.uni-heidelberg.de.

**Figure 1.** (a) Kinesin-1 and kinesin-2 molecular motors, which walk along microtubules for organelle transport.[20] (b) DNA walker moves and transfers cargos on a DNA origami tile.[26] (c) Gold nanorod (yellow) performs stepwise walking on a DNA origami track. During this process, it optically interacts with another gold nanorod (red), leading to circular dichroism spectral changes in real time.[27] (d) Two DNA robots that perform a cargo sorting task on a DNA origami surface, transporting fluorescence molecules to separated destinations.[28]

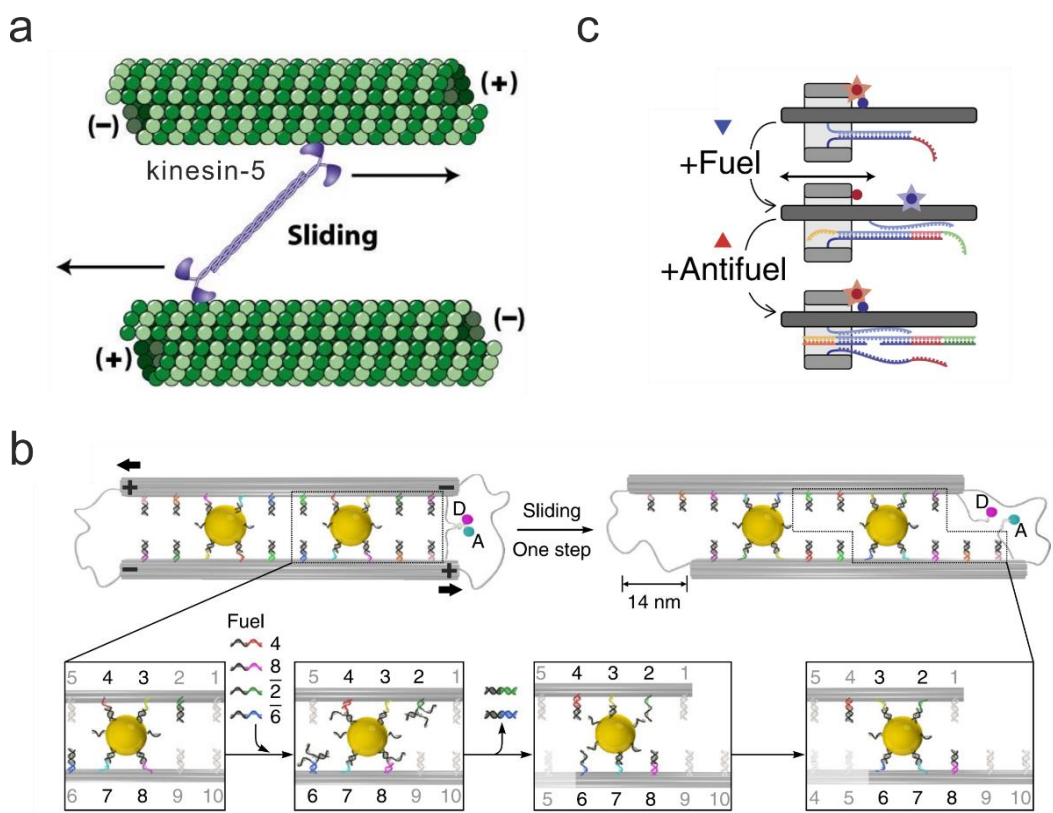

**Figure 2.** (a) Kinesin-5, a bipolar homotetrameric motor protein, uses pairs of dimeric motor domains to crosslink and slide apart antiparallel microtubules.[20] (b) Two gold nanocrystals crosslink and slide two DNA origami filaments in antiparallel.[32] (c) Fuel/anitfuel mechanism used to switch the connection between the two units in a DNA origami rotaxane, enabling long-range sliding movement.[35]

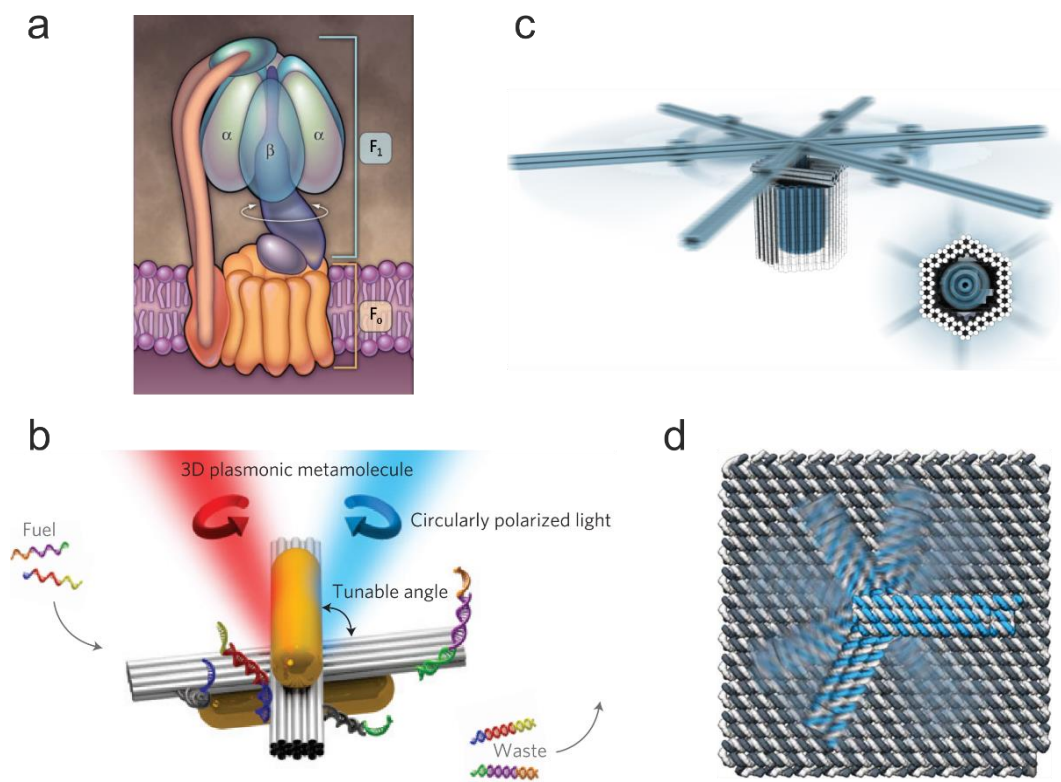

**Figure 3.** (a) Schematic of the $F_0F_1$-ATP synthase complex.[36] (b) Reconfigurable plasmonic metamolecule that is switched between predesignated configurations.[39] (c) Nanoscale rotary apparatus formed from tight-fitting DNA origami components.[40] (d) Self-assembled nanoscale robotic arm controlled by electric fields for rotation.[41]